\documentclass[a4paper,11pt]{article}
\usepackage{pos}

\usepackage{amsmath,amsthm,bbm}
\usepackage[capitalise]{cleveref}

\usepackage{subcaption}
\usepackage{ragged2e}
\DeclareCaptionJustification{justified}{\justifying}
\newcommand{\SU}{\operatorname{SU}}
\renewcommand{\O}{\operatorname{O}}
\newcommand{\SO}{\operatorname{SO}}
\newcommand{\e}{\operatorname{e}}
\renewcommand{\d}{\mathrm{d}}
\renewcommand{\i}{\mathrm{i}}

\newcommand{\<}{\langle}
\renewcommand{\>}{\rangle}
\newcommand{\singlet}{\textnormal{singlet}}
\newcommand{\eff}{\textnormal{eff}}

\usepackage{makecell}

\title{A spin-charge flip symmetric fixed point in 2+1d with massless Dirac fermions}

\author*[1]{Hanqing Liu}
\note{Work done in collaboration with Shailesh Chandrasekharan, Emilie Huffman and Ribhu Kaul.}

\affiliation{Department of Physics, Duke University,\\
  Box 90305, Durham, NC 27708, USA}

\emailAdd{hanqing.liu@duke.edu}

\abstract{We study a quantum phase transition of electrons on a two-dimensional square lattice. Our lattice model preserves the full $\O(4)$ symmetry of free spin-$\frac{1}{2}$ Dirac fermions on a bipartite lattice. In particular, it not only preserves the usual $\SO(4)$ (spin-charge) symmetry like in the half-filling Hubbard model, but also preserves a $\mathbb{Z}_2$ spin-charge flip symmetry. Using sign-problem-free Monte Carlo simulation, we find a second order quantum phase transition from a massless Dirac phase to a massive phase with spontaneously chosen spin order or charge order, which become simultaneously critical at the critical point. We analyze all the possible 4-fermion couplings in the continuum respecting the lattice symmetry, and identify the terms whose effective potential in the broken phase is consistent with the numerical results. Using renormalization group calculations in the continuum, we show the existence of the new spin-charge flip symmetric fixed point and calculate its critical exponents.}

\FullConference{%
 The 38th International Symposium on Lattice Field Theory, LATTICE2021
  26th-30th July, 2021
  Zoom/Gather@Massachusetts Institute of Technology
}

\begin{document}
\maketitle

\section{Introduction}
Understanding the mechanism of mass generation in $2+1$d relativistic fermions is of interest in both condensed matter physics \cite{Ryu:2009,You:2017ltx,Herbut:2006cs} and high energy physics \cite{Rosenstein:1990nm,Zinn-Justin:1991ksq,Ayyar:2016lxq}. One of the most important mechanisms is through spontaneous symmetry breaking driven by strong four-fermion interactions. The study of $2+1$d relativistic fermions also leads to the idea of deconfined quantum criticality \cite{Senthil:2003eed,Assaad:2016flj,Sato:2017tgx,You:2017ltx} and emergent symmetry \cite{Nahum:2015vka,Wang:2017txt,Janssen:2017lqz,Roy:2017vkg,Li:2019acc}.

Due to the non-perturbative nature of this problem, it is important to design lattice models which are amenable to sign-problem-free Monte Carlo simulations. Therefore we study the mass generation in a model of spin-$\frac{1}{2}$ Dirac fermions on a two-dimensional square lattice, which can be simulated efficiently with the fermion bag algorithm \cite{Huffman:2017swn,Huffman:2019efk}. This model is a natural generalization of a $1+1$d model we studied earlier \cite{Liu:2019dvk,Liu:2020ygc}. The model is not only invariant under the $\SO(4)$ symmetry of the Hubbard model at half-filling, but more importantly, it also has an additional $\mathbb{Z}_2$ spin-charge flip symmetry, which combines with the $\SO(4)$ symmetry to form an $\O(4)$ symmetry \cite{Goetz:2021emq}. This $\mathbb{Z}_2$ symmetry protects the renormalization group (RG) flow from leaving the spin-charge flip symmetric subspace, allowing us to explore a new fixed point without fine-tuning. By tuning a single coupling, our model undergoes a quantum phase transition from a massless Dirac fermion phase to a massive phase with either an anti-ferromagnetic (spin) order or a superconducting-CDW (charge), and they become simultaneously critical at the spin-charge flip symmetric fixed point, as confirmed by the Monte Carlo simulation. If we add a Hubbard coupling which breaks the $\mathbb{Z}_2$ symmetry, our model will flow to the usual spin or charge fixed points, which can be described by the ``chiral Heisenberg university class'' \cite{Rosenstein:1993zf,Sorella:2012,Assaad:2013xua,Janssen:2014gea,Classen:2015ssa,Otsuka:2015iba,Zerf:2017zqi,Otsuka:2020lhc,Xu:2020qbj}. This contribution will focus on the continuum analysis of the model, while the numerical results can be found in \cite{Liu:2021otq,Huffman:2021}.

This contribution is organized as follows. In \cref{sec:lattice}, we write down the lattice Hamiltonian and identify its symmetries, and in \cref{sec:continuum}, we map the symmetries of the Hamiltonian to the continuum Lagrangian. Then all the independent four-fermion couplings in the continuum respecting those lattice symmetries are constructed in \cref{sec:interactions}. In \cref{sec:effective-potential}, we identify the relevant interactions whose effective potential in the broken phase is consistent with our numerical results. Finally in \cref{sec:RG}, we calculate the $\beta$ functions, confirm the existence of a new spin-charge flip symmetric fixed point and evaluate the critical exponents in $4-\varepsilon$ dimension.

\section{The lattice Hamiltonian and its symmetries} \label{sec:lattice}
The lattice model we study can be described by the Hamiltonian
\begin{align}
  H &= - \sum_{\< ij \>} \exp \Big(\kappa \eta_{ij} \sum_{\alpha} (c^\dagger_{i\alpha} c_{j\alpha} + c^\dagger_{j\alpha} c_{i\alpha})\Big), \label{eq:Hamiltonian1}
\end{align}
where $\< ij \>$ means $i$ and $j$ are nearest neighbor sites on a square lattice, $\alpha = 1,2$, $\eta_{ij}$ are phases that create the $\pi$-flux, $\kappa$ is the coupling of the model. If we expand the exponent, $H$ can be written in a more conventional form
\begin{align}
H  &\propto -\sum_{\< ij \>}\prod\limits_\alpha \Big[-t\eta_{ij}(c^\dagger_{i\alpha} c_{j\alpha} + c^\dagger_{j\alpha} c_{i\alpha}) + V \Big(n_{i\alpha}-\frac{1}{2}\Big) \Big(n_{j\alpha}-\frac{1}{2}\Big) -\frac{t^2}{V}\Big], \label{eq:Hamiltonian2}
\end{align}
where $V/t = 2\tanh\frac{\kappa}{2}$. The original form of $H$ in \cref{eq:Hamiltonian1} makes it clear that each bond of the Hamiltonian is only a function of the free hopping term, while from \cref{eq:Hamiltonian2} we can see that each bond is a product of the $t-V$ Hamiltonian \cite{Huffman:2017swn,Huffman:2019efk}. If we further expand the terms in \cref{eq:Hamiltonian2} into the quadratic, quartic and higher order terms we get
\thickmuskip=3mu
\begin{align}
H \propto -\sum_{\< ij \>} \Big[ t \eta_{ij} \sum_\alpha (c^\dagger_{i\alpha} c_{j\alpha} + c^\dagger_{j\alpha} c_{i\alpha}) + \frac{V}{2} \Big( \sum_\alpha (c^\dagger_{i\alpha} c_{j\alpha} + c^\dagger_{j\alpha} c_{i\alpha}) \Big)^2 + (\textnormal{6th and 8th orders}) \Big]. \label{eq:Hamiltonian3}
\end{align}
\thickmuskip=5mu
In this form, we can identify the quadratic and quartic terms to the ones in the model studied in \cite{Li:2019acc} with $V = J/2$ in their notation.

Since each bond in this Hamiltonian is exponential of the free hopping term, it has all the space-time and internal symmetries of the free Hamiltonian: spatial translations by one unit $T_a^{1,2}$, $\mathbb{Z}_4$ rotation symmetry $R$, parity $P$, time-reversal $\Theta$ and charge conjugation $C$, an $\SU(2)_s \times \SU(2)_c$ spin-charge symmetry, or equivalently, $\SO(4)$ symmetry, which is manifest in the Majorana Language, and most importantly, a $\mathbb{Z}_2$ spin-charge flip symmetry, or equivalently, charge conjugation on a single layer, which enhances the internal symmetry to $\O(4)$.

From the viewpoint of Wilson RG, all interactions respecting the symmetries of the lattice Hamiltonian can be generated in the continuum. Therefore it is important to know how the symmetries of the Hamiltonian are mapped to the continuum. We can understand this by using the free lattice Hamiltonian which we will do next.

\section{Embedding lattice symmetries in the continuum} \label{sec:continuum}
Let us consider free staggered fermions on a square lattice, given by the first term in \cref{eq:Hamiltonian3}. Linearizing the dispersion relation of this Hamiltonian near the Fermi points, we get the following continuum Hamiltonian
\begin{align}
  H_0 = - \int\d^2x ~& \i \psi_\alpha^{\dagger} (\Gamma^1\partial_1 + \Gamma^2\partial_2) \psi_\alpha,
\label{eq:contH}
\end{align}
where $\Gamma^1 = \sigma^3\otimes\sigma^3$ and $\Gamma^2 = \sigma^3\otimes\sigma^1$ come from the four corners of the Brillouin zone. Using Grassmann coherent fermion path integral, we can rewrite \cref{eq:contH} as the following Euclidean Lagrangian density
\begin{align}
  \mathcal{L}_0 = -\bar\psi_\alpha \gamma^\mu\partial_\mu \psi_\alpha,
\end{align}
where $\psi_\alpha$ is a 4-component Dirac fermion, $\bar\psi_\alpha = \psi_\alpha^\dagger\gamma^0$, $\alpha = 1,2$, and $\mu = 1,2,3$. Here  $\gamma^{0,1,2,3,5}$ are five $4\times 4$ Hermitian matrices satisfying the Clifford algebra $\{ \gamma^i, \gamma^j \} = 2 \delta^{ij} \mathbbm{1}_4$. We can choose $\gamma^{0,3,5}$ to be real and $\gamma^{1,2}$ to be imaginary. One basis consistent with $\Gamma^{1,2}$ in \cref{eq:contH} is given by
\begin{align}
  \gamma^0 = \sigma^1 \otimes \mathbbm{1}, ~ \gamma^1 = \sigma^2 \otimes \sigma^3, ~ \gamma^2 = \sigma^2 \otimes \sigma^1, ~ \gamma^3 = \sigma^3 \otimes \mathbbm{1}, ~ \gamma^5 = \gamma^0\gamma^1\gamma^2\gamma^3 = \sigma^2\otimes\sigma^2 . 
\end{align}
Space-time transformations on the lattice mix Dirac components in the continuum as follows
\begin{align}
  T_a^{1,2}: \psi \mapsto \i\gamma^{3,5}\psi, ~ R: \psi \mapsto \e^{\i\frac{\pi}{4}(\i\gamma^1\gamma^2 + \i\gamma^3\gamma^5)} \psi, ~ P: \psi \mapsto \i\gamma^5\gamma^1\psi, ~ \Theta: \psi \mapsto \gamma^0K\psi,
\end{align}
where 
$K$ is the complex conjugation operator. Except for $R$, all of them act as $\mathbb{Z}_2$ symmetries on fermion bilinears. $R$ acts as a $\mathbb{Z}_4$ symmetry on the lattice, and will be enhanced to an $\SO(2)_R$ internal symmetry 
\begin{align}
  \tilde R: \psi \mapsto \e^{\i\frac{\theta}{2}( \i\gamma^3\gamma^5)} \psi
\end{align}
in the continuum. When analyzing the internal symmetries, especially the charge symmetry, it is more convenient to use Majorana representation $\psi_1 = \xi_1 - \i \xi_2$, $\psi_2 = \xi_3 - \i \xi_4$, and
\begin{align}
  \mathcal{L}_0 = -\xi_a^T \gamma^0\gamma^\mu\partial_\mu \xi_a,
  \label{eq:majoranaL}
\end{align}
where $a = 1,2,3,4$, and each $\xi_a$ is a four-component Majorana fermion. Clearly \cref{eq:majoranaL} has an $\O(4)$ symmetry, which is nothing but the spin-charge symmetry $\SO(3)_s\times \SO(3)_c$ and the spin-charge flip symmetry $\mathbb{Z}_2^{sc}$ on the lattice. In fact, it turns out that this continuum Lagrangian actually has an $\O(8)$ symmetry. Thus the lattice realizes the subgroup $\SO(3)_s \times \SO(3)_c \times \mathbb{Z}_2^{sc} \times \SO(2)_R$ of this $\O(8)$ symmetry in the continuum.

\section{Interactions respecting the lattice symmetries} \label{sec:interactions}
In this section, we analyze all the Lorentz-invariant interactions allowed by the lattice symmetries. First, the allowed four-fermion interactions must be singlets under $\SO(3)_s \times \SO(3)_c \times \SO(2)_R$, and they can be constructed from fermion bilinears, including space-time (pseudo-)scalars, i.e., masses, and space-time (pseudo-)vectors, i.e., currents. The fermion bilinears form reducible representations of the symmetry group $\SO(3)_s \times \SO(3)_c \times \SO(2)_R$, and decompose into irreducible representations (irreps) as 36 masses,
\begin{align}
  \mathbf{36} = (\mathbf{3} \otimes \mathbf{1} + \mathbf{1} \otimes \mathbf{3}) \otimes \mathbf{1} + (\mathbf{3} \otimes \mathbf{3} + \mathbf{1} \otimes \mathbf{1}) \otimes (\mathbf{2} + \mathbf{1}),
\label{eq:massterms}
\end{align}
which agrees with a previous work \cite{Ryu:2009}, and 28 currents,
\begin{align}
  \mathbf{28} = (\mathbf{3} \otimes \mathbf{1} + \mathbf{1} \otimes \mathbf{3}) \otimes (\mathbf{2} + \mathbf{1}) + (\mathbf{3} \otimes \mathbf{3} + \mathbf{1} \otimes \mathbf{1}) \otimes \mathbf{1}.
\label{eq:currterms}
\end{align}
If we also take into account the $\mathbb{Z}_2$ lattice space-time symmetries, no mass terms are invariant under all of them, and therefore our continuum theory cannot have any mass terms. Building singlets from bilinear irreps that transform according to \cref{eq:massterms,eq:currterms}, we get 6 Gross-Neveu couplings and 6 Thirring couplings, and these $\mathbb{Z}_2$ lattice symmetries are automatically satisfied, while the spin-charge flip symmetry $\mathbb{Z}_2^{sc}$ flips some of those terms. However, due to the Fierz identity, only $4$ of these $12$ couplings are independent, and remarkably, they can all be chosen to be Gross-Neveu couplings,
\begin{align}
  &\mathcal{L}_s = \frac{g_s^2}{2} |\vec M_s|^2, \quad \mathcal{L}_c = \frac{g_c^2}{2} |\vec M_c|^2, \quad \mathcal{L}_R = \frac{g_R^2}{2} |\vec M_R|^2, \quad \mathcal{L}_\singlet = \frac{g_\singlet^2}{2} M_\singlet^2,
\end{align}
where
\begin{align}
  \vec M_s &= \bar \psi_\alpha \vec \sigma_{\alpha\beta} \psi_\beta , \quad \vec M_c = (\psi^{T}_2\gamma^0 \psi_1+\bar\psi_1\gamma^0\bar\psi^{T}_2 ,  \i( \psi^{T}_2\gamma^0 \psi_1-\bar\psi_1\gamma^0\bar\psi^{T}_2 ), \bar\psi_1 \psi_1 + \bar\psi_2 \psi_2), \nonumber\\
  \vec M_R &= (\bar\psi_\alpha \i \gamma^3 \psi_\alpha, \bar\psi_\alpha \i \gamma^5 \psi_\alpha), \quad M_\singlet = \bar\psi_\alpha \i \gamma^3\gamma^5 \psi_\alpha.
\end{align}
For example, in \cite{Li:2019acc}, the authors use $\mathcal{L}_R$ built from $\mathbf{1} \otimes \mathbf{1} \otimes \mathbf{2}$ to study a phase transition from Dirac phase to Kekul\'e valance-bond-solid (VBS) phase.

\section{The continuum model and the effective potential} \label{sec:effective-potential}

From our Monte Carlo results \cite{Liu:2021otq,Huffman:2021}, we see either anti-ferromagnetic (spin) order or superconducting-CDW (charge) order at strong couplings, but no VBS order, i.e., $|\< \vec M_s \>| \neq 0$ or  $|\< \vec M_c \>| \neq 0$, but $|\< \vec M_R \>| = 0$. Therefore we believe our model represents a lattice regularization of the Gross-Neveu model with spin and charge couplings given by the Lagrangian density
\begin{align}\label{eq:Lagrangian}
  \mathcal{L}_{\rm GN} = \mathcal{L}_0 + \mathcal{L}_s + \mathcal{L}_c.
\end{align}
Furthermore, the $\mathbb{Z}_2^{sc}$ spin-charge flip symmetry of the lattice model imposes the restriction that $g_s^2 = g_c^2$. Adding interactions to the lattice model that breaks the spin-charge flip symmetry, like the Hubbard coupling, would lead to $g_s^2 \neq g_c^2$.

We can confirm the expected symmetry breaking pattern of the above interaction by calculating the one-loop effective potential. In order to do so, we introduce auxiliary scalar fields $\vec\phi_s$ and $\vec\phi_c$ which transform in the $\mathbf{3} \otimes \mathbf{1} \otimes \mathbf{1}$ and $\mathbf{1} \otimes \mathbf{3} \otimes \mathbf{1}$ representations respectively, and rewrite the Lagrangian as,
\begin{align}
\mathcal{L}^{\textnormal{aux}}_{\textnormal{GN}} &= -\bar\psi_\alpha \gamma^\mu\partial_\mu \psi_\alpha + \frac{1}{2g_s^2}\phi_s^2 + \frac{1}{2g_c^2}\phi_c^2 + \phi_s(\bar\psi^1 \psi^1 - \bar\psi^2 \psi^2) + \phi_c(\bar\psi^1 \psi^1 + \bar\psi^2 \psi^2),
\end{align}
where we have rotated $\vec\phi_s$ and $\vec\phi_c$ such that $\phi_s^{1,2} \equiv \phi_c^{1,2} \equiv 0$, and relabeled $\phi_s^3 \rightarrow \phi_s$ and $\phi_c^3 \rightarrow \phi_c$, because the boson fields will be treated as constants in space-time. By integrating out the quadratic fermions, we get the following effective potential of the $\phi_s$ and $\phi_c$ fields \cite{Liu:2021otq},
\begin{align}
  \frac{1}{S_d}&V_\eff[\phi_s, \phi_c] \nonumber \\
  &= \frac{2\pi}{3}|\phi_c-\phi_s|^3 -\frac{4}{3}(\phi_c-\phi_s)^3\tan^{-1}\frac{\phi_c-\phi_s}{\Lambda} +\frac{2\pi}{3}|\phi_c+\phi_s|^3 -\frac{4}{3}(\phi_c+\phi_s)^3\tan^{-1}\frac{\phi_c+\phi_s}{\Lambda} \nonumber\\
  &\quad - \frac{8}{3}\Lambda(\phi_c^2 + \phi_s^2) - \frac{2}{3}\Lambda^3\log\Big(1 + 2\frac{\phi_c^2 + \phi_s^2}{\Lambda^2} + \frac{(\phi_c^2 - \phi_s^2)^2}{\Lambda^4}\Big) + \frac{1}{2S_dg_s^2}\phi_s^2 + \frac{1}{2S_dg_c^2}\phi_c^2,
\end{align}
where $S_d = \frac{2}{(4\pi)^{d/2}\Gamma(d/2)}$ is the loop integral factor and here we should set $d = 3$, and $\Lambda$ is the cutoff. 

\begin{figure}[htp]
  \centering
  \includegraphics[width=0.5\textwidth]{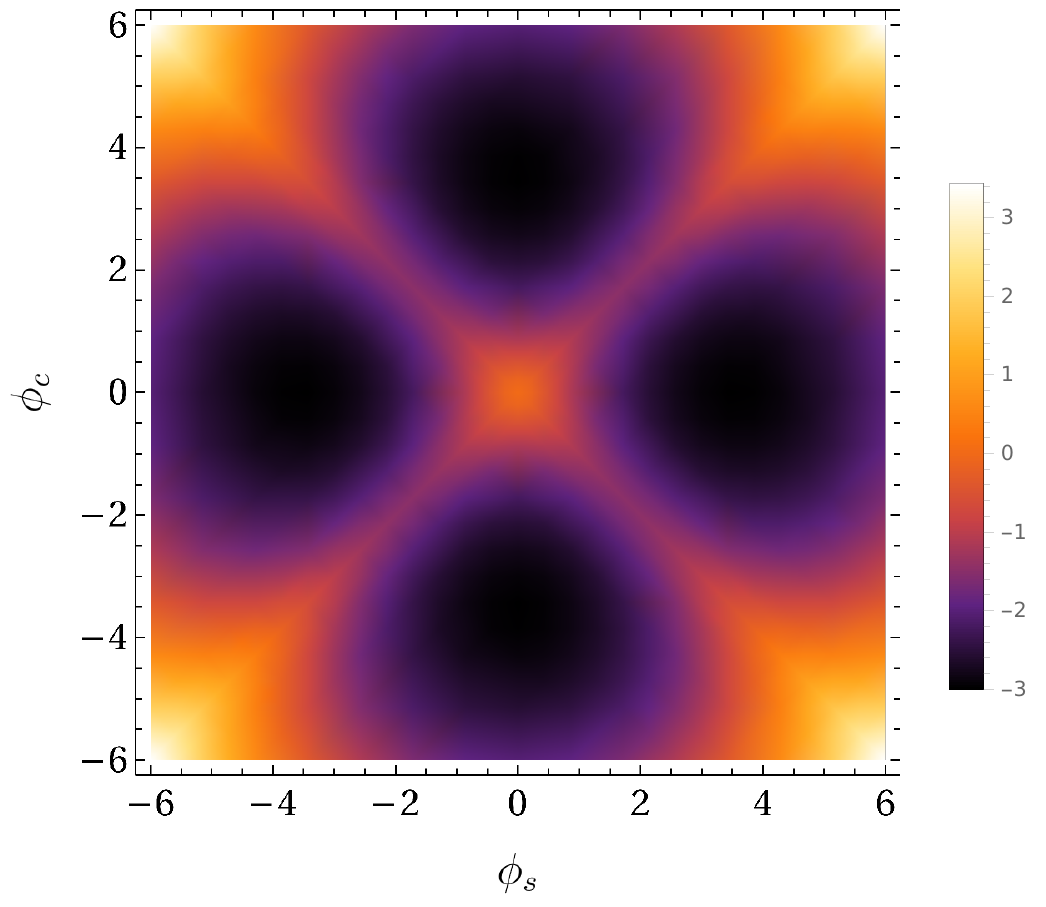}
  \caption{Effective potential in the broken phase at $S_d\Lambda g_s^2 = S_d\Lambda g_c^2 = 5$.}
  \label{fig:effective-potential}
\end{figure}

In \cref{fig:effective-potential} we plot the effective potential in the broken phase at $S_d\Lambda g_s^2 = S_d\Lambda g_c^2 = 5$. The critical values of $g_s$ and $g_c$ are at $S_d\Lambda g_s^2 = S_d\Lambda g_c^2 = \frac{1}{8}$. From the effective potential we see that in the broken phase, we have either spin order or charge order, but not both, and therefore the $\mathbb{Z}_2^{sc}$ spin-charge flip symmetry is also spontaneously broken.

\section{RG analysis and critical exponents} \label{sec:RG}
In order to understand the RG flow and critical properties of the Lagrangian in \cref{eq:Lagrangian}, we calculate the usual $4-\varepsilon$ expansion using the corresponding Gross-Neveu-Yukawa Lagrangian
\begin{align}
  \mathcal{L}_{\textnormal{GNY}} &= - \bar \psi_\alpha \gamma^\mu \partial_\mu \psi_\alpha + g_s \vec \phi_s \cdot \vec M_s + g_c \vec \phi_c \cdot \vec M_c \nonumber\\
  &+ \sum_{a=s,c}\left ( \frac{1}{2} \partial_\mu \vec\phi_a \cdot \partial^\mu \vec\phi_a + \frac{1}{2} m_a^2 \vec\phi_a \cdot \vec\phi_a + \frac{1}{4!} \lambda_a (\vec\phi_a \cdot \vec\phi_a)^2\right )  + \frac{1}{12} \lambda_{sc} (\vec\phi_s \cdot \vec\phi_s) (\vec\phi_c \cdot \vec\phi_c) .
\end{align}
We have calculated the one-loop $\beta$ functions for the coupling $g_s^2$, $g_c^2$, $\lambda_s$, $\lambda_c$ and $\lambda_{sc}$, and they are given by \cite{Liu:2021otq}
\begin{align}
  \frac{\d g_s^2}{\d\log\mu} &= -\varepsilon g_s^2 + S_d \big( (2N_f+1)g_s^4 + 9 g_s^2g_c^2 \big), \label{eq:RG-g}\\
  \frac{\d \lambda_s}{\d\log\mu} &= -\varepsilon \lambda_s + S_d \Big( \frac{11}{6} \lambda_s^2 + \frac{1}{2} \lambda_{sc}^2 + 4N_fg_s^2\lambda_s - 24N_fg_s^4 \Big), \\
  \frac{\d \lambda_{sc}}{\d\log\mu} &= -\varepsilon\lambda_{sc} + S_d \Big( \frac{5}{6} (\lambda_s + \lambda_c) \lambda_{sc} + \frac{2}{3} \lambda_{sc}^2 + 2N_f(g_s^2 + g_c^2)\lambda_{sc} - 24N_f g_s^2g_c^2 \Big),
\end{align}
where the $\beta$ functions for $g_c^2$ and $\lambda_c$ can be obtained by the spin-charge flip symmetry. 

\begin{figure}[htp]
  \centering
  \begin{subfigure}[b]{0.4\textwidth}
    \includegraphics[width=\textwidth]{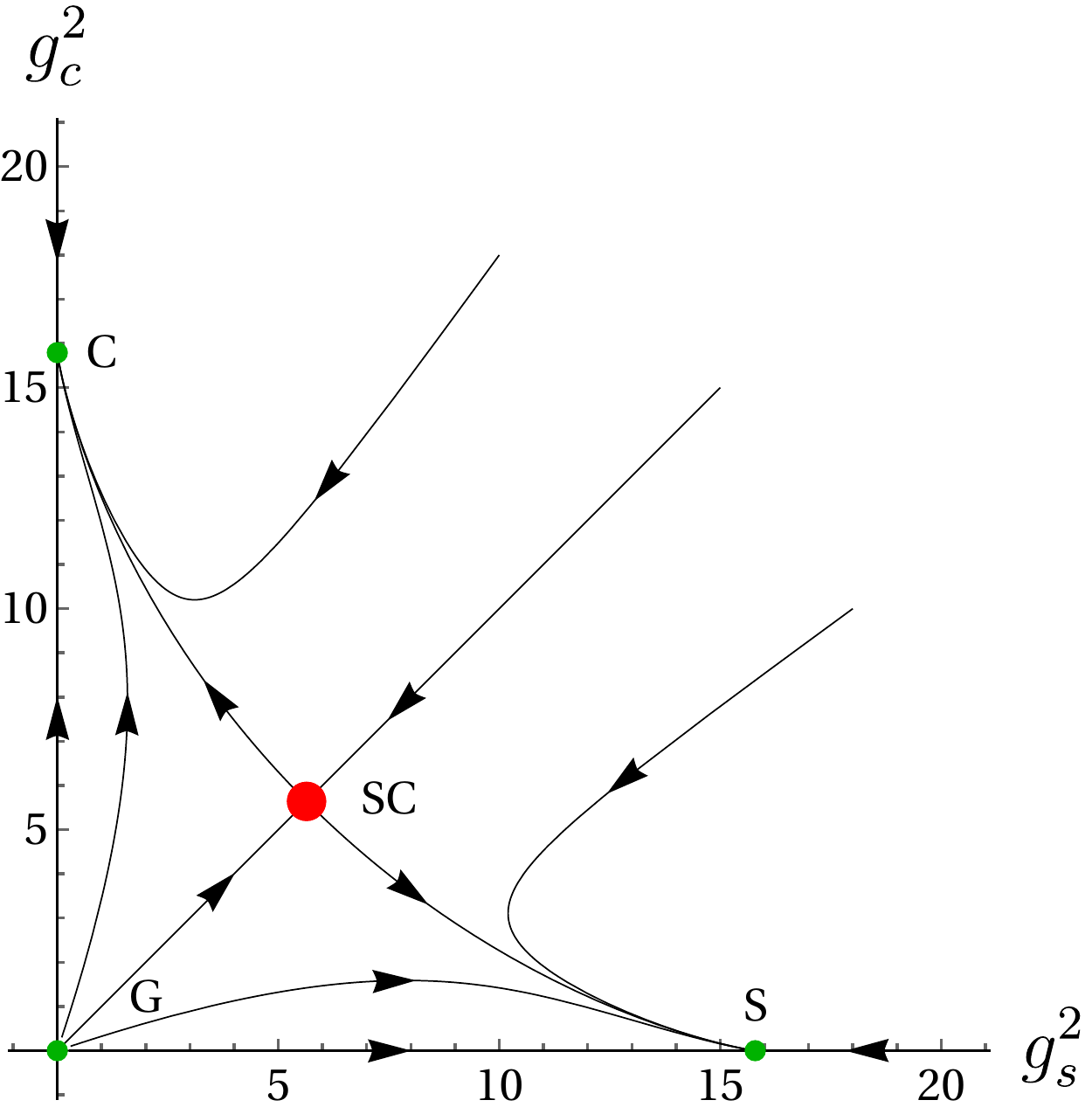}
    \caption{Yukawa couplings}
    \label{fig:RG-g}
  \end{subfigure}
  \begin{subfigure}[b]{0.3\textwidth}
    \includegraphics[width=\textwidth]{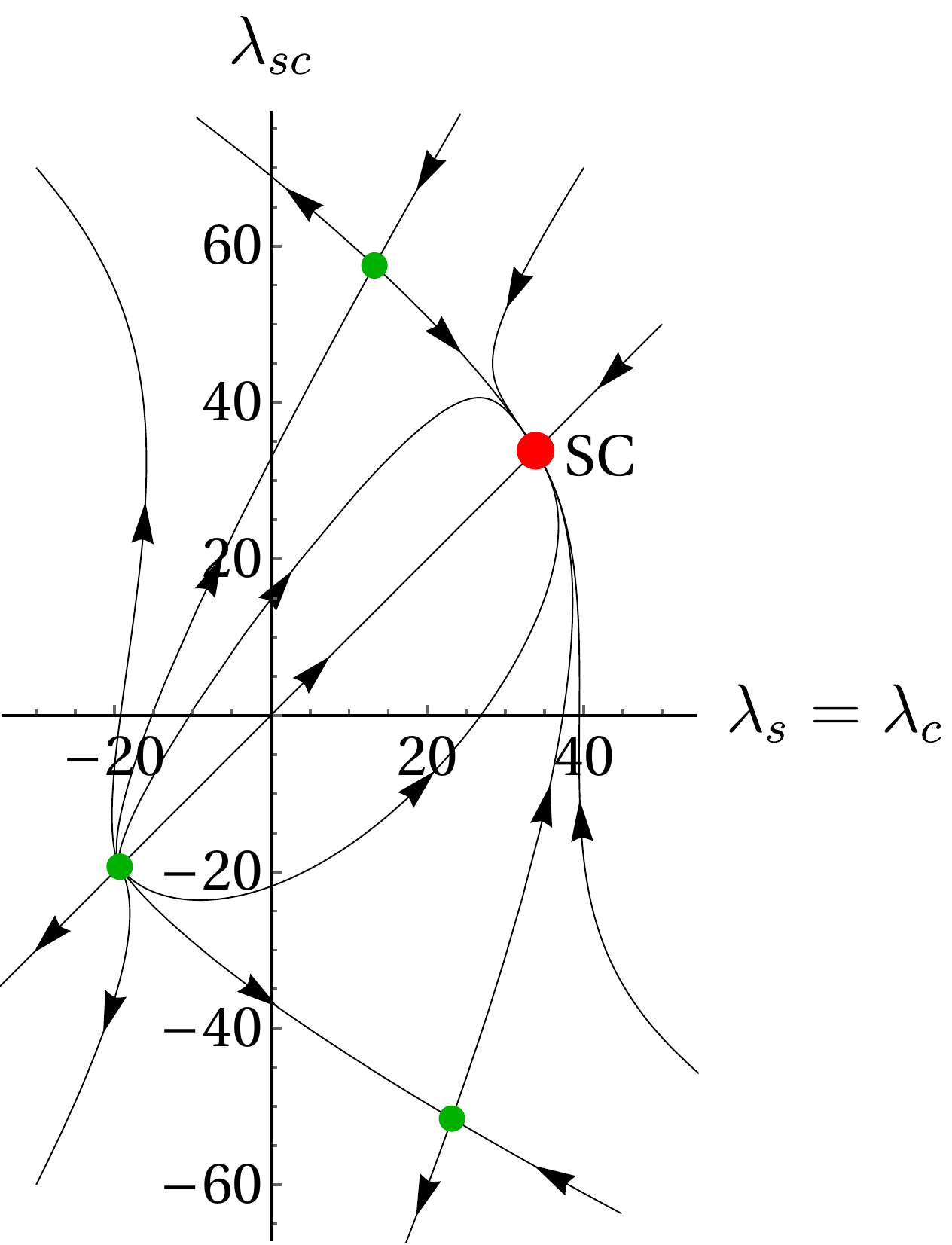}
    \caption{Boson self-interactions}
    \label{fig:RG-lambda}
  \end{subfigure}
  \caption{RG flow of the couplings in $4-\varepsilon$ expansion}
  \label{fig:RG-flow}
\end{figure}

Using these $\beta$ functions, we plot the RG flows in \cref{fig:RG-flow}. From \cref{eq:RG-g} we see that the Yukawa couplings mix only among themselves, and from \cref{fig:RG-g} we see that there is a spin-charge flip symmetric fixed point (SC) on the $g_s^2 = g_c^2$ axis, which separates the massless Dirac phase from the broken phase. The $\mathbb{Z}_2^{sc}$ spin-charge flip symmetry prevents the flow from leaving the diagonal $g_s^2 = g_c^2$ axis. Breaking this $\mathbb{Z}_2^{sc}$ symmetry through the Hubbard interaction for example, would drive the flow away from the diagonal axis to the usual spin or charge fixed points, depending on the sign of the Hubbard coupling. Assuming $g_s^2 = g_c^2$ are at SC, the flow diagram of the boson self-interactions in the spin-charge flip symmetric slice is shown in \cref{fig:RG-lambda}. There is only one stable fixed point, which can be identified as the SC fixed point.

Evaluating the critical exponents at the spin-charge symmetric fixed point using the $4-\varepsilon$ expansion, we have \cite{Liu:2021otq},
\begin{align}
  \eta = \frac{2}{7}\varepsilon, \quad \eta_\psi = \frac{3}{14}\varepsilon, \quad 1/\nu = 2 - \frac{6}{7}\varepsilon.
\end{align}
In the large $N_f$ limit ($N_f$ is the number of values $\alpha$ takes, and in our model $N_f = 2$), we obtain
\begin{align}
  \eta = \varepsilon - \frac{5\varepsilon}{N_f +5}, \quad \eta_\psi = \frac{3}{2(N_f+5)}\varepsilon, \quad 1/\nu = 2 - \varepsilon - \frac{3\varepsilon}{N_f +5}.
\end{align}
These exponents are different from those obtained from the Hubbard model \cite{Rosenstein:1993zf}.

\section{Conclusions}
Our lattice Hamiltonian, which can be solved using the fermion bag approach, has a spin-charge flip symmetry. We have shown that the presence of this additional symmetry leads to a new fixed point that can be reached by tuning a single coupling on the lattice. The fixed point thus describes an interesting phase transition between a massless Dirac fermion phase and a phase featuring spontaneous spin symmetry breaking or charge symmetry breaking, as well as spontaneous spin-charge flip symmetry breaking. Here we uncover the physics of the continuum model by calculating its effective potential and computing the critical exponents using the $4-\varepsilon$ expansion up to one loop.

\acknowledgments
This work was done in collaboration with Shailesh Chandrasekharan, Emilie Huffman and Ribhu Kaul.
The material presented here was supported by the U.S. Department of Energy, Office of Science, Nuclear Physics program under Award Numbers DE-FG02-05ER41368.

\bibliographystyle{JHEP}
\bibliography{Refs,MyPubs,Refs-3dfermion}

\providecommand{\href}[2]{#2}\begingroup\raggedright\begin{thebibliography}{10}

\bibitem{Ryu:2009}
S.~Ryu, C.~Mudry, C.-Y. Hou and C.~Chamon, \emph{Masses in graphenelike
  two-dimensional electronic systems: Topological defects in order parameters
  and their fractional exchange statistics},
  \href{http://dx.doi.org/10.1103/physrevb.80.205319}{\emph{Physical Review B}
  {\bf 80} (Nov, 2009) }.

\bibitem{You:2017ltx}
Y.-Z. You, Y.-C. He, C.~Xu and A.~Vishwanath, \emph{{Symmetric Fermion Mass
  Generation as Deconfined Quantum Criticality}},
  \href{http://dx.doi.org/10.1103/PhysRevX.8.011026}{\emph{Phys. Rev. X} {\bf
  8} (2018) 011026}, [\href{http://arxiv.org/abs/1705.09313}{{\tt
  1705.09313}}].

\bibitem{Herbut:2006cs}
I.~F. Herbut, \emph{{Interactions and phase transitions on graphene's honeycomb
  lattice}}, \href{http://dx.doi.org/10.1103/PhysRevLett.97.146401}{\emph{Phys.
  Rev. Lett.} {\bf 97} (2006) 146401},
  [\href{http://arxiv.org/abs/cond-mat/0606195}{{\tt cond-mat/0606195}}].

\bibitem{Rosenstein:1990nm}
B.~Rosenstein, B.~Warr and S.~H. Park, \emph{{Dynamical symmetry breaking in
  four Fermi interaction models}},
  \href{http://dx.doi.org/10.1016/0370-1573(91)90129-A}{\emph{Phys. Rept.} {\bf
  205} (1991) 59--108}.

\bibitem{Zinn-Justin:1991ksq}
J.~Zinn-Justin, \emph{{Four fermion interaction near four-dimensions}},
  \href{http://dx.doi.org/10.1016/0550-3213(91)90043-W}{\emph{Nucl. Phys. B}
  {\bf 367} (1991) 105--122}.

\bibitem{Ayyar:2016lxq}
V.~Ayyar and S.~Chandrasekharan, \emph{{Fermion masses through four-fermion
  condensates}}, \href{http://dx.doi.org/10.1007/JHEP10(2016)058}{\emph{JHEP}
  {\bf 10} (2016) 058}, [\href{http://arxiv.org/abs/1606.06312}{{\tt
  1606.06312}}].

\bibitem{Senthil:2003eed}
T.~Senthil, A.~Vishwanath, L.~Balents, S.~Sachdev and M.~P.~A. Fisher,
  \emph{{Deconfined Quantum Critical Points}},
  \href{http://dx.doi.org/10.1126/science.1091806}{\emph{Science} {\bf 303}
  (2004) 1490--1494}, [\href{http://arxiv.org/abs/cond-mat/0311326}{{\tt
  cond-mat/0311326}}].

\bibitem{Assaad:2016flj}
F.~F. Assaad and T.~Grover, \emph{{Simple Fermionic Model of Deconfined Phases
  and Phase Transitions}},
  \href{http://dx.doi.org/10.1103/PhysRevX.6.041049}{\emph{Phys. Rev. X} {\bf
  6} (2016) 041049}, [\href{http://arxiv.org/abs/1607.03912}{{\tt
  1607.03912}}].

\bibitem{Sato:2017tgx}
T.~Sato, M.~Hohenadler and F.~F. Assaad, \emph{{Dirac Fermions with Competing
  Orders: Non-Landau Transition with Emergent Symmetry}},
  \href{http://dx.doi.org/10.1103/PhysRevLett.119.197203}{\emph{Phys. Rev.
  Lett.} {\bf 119} (2017) 197203}, [\href{http://arxiv.org/abs/1707.03027}{{\tt
  1707.03027}}].

\bibitem{Nahum:2015vka}
A.~Nahum, P.~Serna, J.~T. Chalker, M.~Ortu\~no and A.~M. Somoza,
  \emph{{Emergent SO(5) Symmetry at the N\'eel to Valence-Bond-Solid
  Transition}},
  \href{http://dx.doi.org/10.1103/PhysRevLett.115.267203}{\emph{Phys. Rev.
  Lett.} {\bf 115} (2015) 267203}, [\href{http://arxiv.org/abs/1508.06668}{{\tt
  1508.06668}}].

\bibitem{Wang:2017txt}
C.~Wang, A.~Nahum, M.~A. Metlitski, C.~Xu and T.~Senthil, \emph{{Deconfined
  quantum critical points: symmetries and dualities}},
  \href{http://dx.doi.org/10.1103/PhysRevX.7.031051}{\emph{Phys. Rev. X} {\bf
  7} (2017) 031051}, [\href{http://arxiv.org/abs/1703.02426}{{\tt
  1703.02426}}].

\bibitem{Janssen:2017lqz}
L.~Janssen, I.~F. Herbut and M.~M. Scherer, \emph{{Compatible orders and
  fermion-induced emergent symmetry in Dirac systems}},
  \href{http://dx.doi.org/10.1103/PhysRevB.97.041117}{\emph{Phys. Rev. B} {\bf
  97} (2018) 041117}, [\href{http://arxiv.org/abs/1711.11042}{{\tt
  1711.11042}}].

\bibitem{Roy:2017vkg}
B.~Roy, P.~Goswami and V.~Juricic, \emph{{Itinerant quantum multicriticality of
  two-dimensional Dirac fermions}},
  \href{http://dx.doi.org/10.1103/PhysRevB.97.205117}{\emph{Phys. Rev. B} {\bf
  97} (2018) 205117}, [\href{http://arxiv.org/abs/1712.05400}{{\tt
  1712.05400}}].

\bibitem{Li:2019acc}
Z.-X. Li, S.-K. Jian and H.~Yao, \emph{{Deconfined quantum criticality and
  emergent SO(5) symmetry in fermionic systems}},
  \href{http://arxiv.org/abs/1904.10975}{{\tt 1904.10975}}.

\bibitem{Huffman:2017swn}
E.~Huffman and S.~Chandrasekharan, \emph{{Fermion bag approach to Hamiltonian
  lattice field theories in continuous time}},
  \href{http://dx.doi.org/10.1103/PhysRevD.96.114502}{\emph{Phys. Rev. D} {\bf
  96} (2017) 114502}, [\href{http://arxiv.org/abs/1709.03578}{{\tt
  1709.03578}}].

\bibitem{Huffman:2019efk}
E.~Huffman and S.~Chandrasekharan, \emph{{Fermion-bag inspired Hamiltonian
  lattice field theory for fermionic quantum criticality}},
  \href{http://dx.doi.org/10.1103/PhysRevD.101.074501}{\emph{Phys. Rev. D} {\bf
  101} (2020) 074501}, [\href{http://arxiv.org/abs/1912.12823}{{\tt
  1912.12823}}].

\bibitem{Liu:2019dvk}
H.~Liu, \emph{{Quantum Critical Phenomena in an $O(4)$ Fermion Chain}},  in
  \emph{{37th International Symposium on Lattice Field Theory (Lattice 2019)
  Wuhan, Hubei, China, June 16-22, 2019}}, 2019.
\newblock \href{http://arxiv.org/abs/1912.11237}{{\tt 1912.11237}}.

\bibitem{Liu:2020ygc}
H.~Liu, S.~Chandrasekharan and R.~K. Kaul, \emph{{Hamiltonian models of lattice
  fermions solvable by the meron-cluster algorithm}},
  \href{http://dx.doi.org/10.1103/PhysRevD.103.054033}{\emph{Phys. Rev. D} {\bf
  103} (2021) 054033}, [\href{http://arxiv.org/abs/2011.13208}{{\tt
  2011.13208}}].

\bibitem{Goetz:2021emq}
A.~Goetz, S.~Beyl, M.~Hohenadler and F.~F. Assaad, \emph{{Langevin dynamics
  simulations of the two-dimensional Su-Schrieffer-Heeger model}},
  \href{http://arxiv.org/abs/2102.08899}{{\tt 2102.08899}}.

\bibitem{Rosenstein:1993zf}
B.~Rosenstein, H.-L. Yu and A.~Kovner, \emph{{Critical exponents of new
  universality classes}},
  \href{http://dx.doi.org/10.1016/0370-2693(93)91253-J}{\emph{Phys. Lett. B}
  {\bf 314} (1993) 381--386}.

\bibitem{Sorella:2012}
S.~Sorella, Y.~Otsuka and S.~Yunoki, \emph{Absence of a spin liquid phase in
  the hubbard model on the honeycomb lattice},
  \href{http://dx.doi.org/10.1038/srep00992}{\emph{Scientific Reports} {\bf 2}
  (Dec, 2012) }.

\bibitem{Assaad:2013xua}
F.~F. Assaad and I.~F. Herbut, \emph{{Pinning the order: the nature of quantum
  criticality in the Hubbard model on honeycomb lattice}},
  \href{http://dx.doi.org/10.1103/PhysRevX.3.031010}{\emph{Phys. Rev. X} {\bf
  3} (2013) 031010}, [\href{http://arxiv.org/abs/1304.6340}{{\tt 1304.6340}}].

\bibitem{Janssen:2014gea}
L.~Janssen and I.~F. Herbut, \emph{{Antiferromagnetic critical point on
  graphene's honeycomb lattice: A functional renormalization group approach}},
  \href{http://dx.doi.org/10.1103/PhysRevB.89.205403}{\emph{Phys. Rev. B} {\bf
  89} (2014) 205403}, [\href{http://arxiv.org/abs/1402.6277}{{\tt 1402.6277}}].

\bibitem{Classen:2015ssa}
L.~Classen, I.~F. Herbut, L.~Janssen and M.~M. Scherer, \emph{{Mott
  multicriticality of Dirac electrons in graphene}},
  \href{http://dx.doi.org/10.1103/PhysRevB.92.035429}{\emph{Phys. Rev. B} {\bf
  92} (2015) 035429}, [\href{http://arxiv.org/abs/1503.05002}{{\tt
  1503.05002}}].

\bibitem{Otsuka:2015iba}
Y.~Otsuka, S.~Yunoki and S.~Sorella, \emph{{Universal quantum criticality in
  the metal-insulator transition of two-dimensional interacting Dirac
  electrons}}, \href{http://dx.doi.org/10.1103/PhysRevX.6.011029}{\emph{Phys.
  Rev. X} {\bf 6} (2016) 011029}, [\href{http://arxiv.org/abs/1510.08593}{{\tt
  1510.08593}}].

\bibitem{Zerf:2017zqi}
N.~Zerf, L.~N. Mihaila, P.~Marquard, I.~F. Herbut and M.~M. Scherer,
  \emph{{Four-loop critical exponents for the Gross-Neveu-Yukawa models}},
  \href{http://dx.doi.org/10.1103/PhysRevD.96.096010}{\emph{Phys. Rev. D} {\bf
  96} (2017) 096010}, [\href{http://arxiv.org/abs/1709.05057}{{\tt
  1709.05057}}].

\bibitem{Otsuka:2020lhc}
Y.~Otsuka, K.~Seki, S.~Sorella and S.~Yunoki, \emph{{Dirac electrons in the
  square-lattice Hubbard model with a $d$-wave pairing field: The chiral
  Heisenberg universality class revisited}},
  \href{http://dx.doi.org/10.1103/PhysRevB.102.235105}{\emph{Phys. Rev. B} {\bf
  102} (2020) 235105}, [\href{http://arxiv.org/abs/2009.04685}{{\tt
  2009.04685}}].

\bibitem{Xu:2020qbj}
X.~Y. Xu and T.~Grover, \emph{{Competing Nodal $d$-Wave Superconductivity and
  Antiferromagnetism}},
  \href{http://dx.doi.org/10.1103/PhysRevLett.126.217002}{\emph{Phys. Rev.
  Lett.} {\bf 126} (2021) 217002}, [\href{http://arxiv.org/abs/2009.06644}{{\tt
  2009.06644}}].

\bibitem{Liu:2021otq}
H.~Liu, E.~Huffman, S.~Chandrasekharan and R.~K. Kaul, \emph{{Quantum
  Criticality of Anti-ferromagnetism and Superconductivity with Relativity}},
  \href{http://arxiv.org/abs/2109.06059}{{\tt 2109.06059}}.

\bibitem{Huffman:2021}
E.~Huffman, \emph{{Proceeding of the 38th International Symposium on Lattice
  Field Theory, LATTICE2021}}, .

\end{thebibliography}\endgroup



\end{document}